\documentstyle[aps,prb,epsfig,multicol]{revtex}

\begin{document}
\draft

\title{Current-induced highly dissipative domains in high $T_c$ thin films}
\author{S.~Reymond\,\cite{email}, L.~Antognazza, M.~Decroux,
E.~Koller, P.~Reinert and \O .~Fischer}
\address{DPMC, Universit\'e de Gen\`eve, Quai Ernest-Ansermet 24,
1211 Gen\`eve 4, Switzerland}
\date{\today}

\maketitle

\begin{abstract}

We have investigated the resistive response of high $T_c$ thin
films submitted to a high density of current. For this purpose,
current pulses were applied into bridges made of
Nd$_{1.15}$Ba$_{1.85}$Cu$_3$O$_{7-\delta}$ and
Bi$_2$Sr$_2$CaCu$_2$O$_{8+\delta}$. By recording the time
dependent voltage, we observe that at a certain critical current
$j^*$, a highly dissipative domain develops somewhere along the
bridge. The successive formation of these domains produces stepped
I-V characteristics. We present evidences that these domains are
not regions with a temperature above $T_c$, as for hot spots. In
fact this phenomenon appears to be analog to the nucleation of
phase-slip centers observed in conventional superconductors near
$T_c$, but here in contrast they appear in a wide temperature
range. Under some conditions, these domains will propagate and
destroy the superconductivity within the whole sample. We have
measured the temperature dependence of $j^*$ and found a similar
behavior in the two investigated compounds. This temperature
dependence is just the one expected for the depairing current, but
the amplitude is about 100 times smaller.

\end{abstract}


\begin{multicols}{2}


\section{introduction}


It has been observed in the early 70's that in one-dimensional
(1D) classical superconductors quantum effects can dominate the
dissipation and give rise to spectacular phenomena such as voltage
steps in the I-V curves~\cite{Web2,Mey1}. The steps are due to the
sequential development of phase-slip centers (PSC)~\cite{Sko1},
which are dissipative regions, where the superconducting order
parameter oscillates. Each time it drops to zero the phase of the
order parameter 'slips' by 2$\pi$ \ \cite{Not1}. The PSC formation
could be observed only in a very narrow interval of temperature
below $T_c$ \ \cite{T_inter}. At lower temperature, sharp features
in the I-V characteristics were also observed, but caused by a
different mechanism: thermal instabilities resulting in the
development of localized normal regions where the temperature
exceeds $T_c$ (hot spots)~\cite{Gur1,Sko2}.

Do high $T_c$ compounds behave similarly? Signatures of phase
slippage as been reported near $T_c$ in
YBa$_2$Cu$_3$O$_{7-\delta}$ grain boundary junctions~\cite{Gro1},
but also at low temperature in YBa$_2$Cu$_3$O$_{7-\delta}$
films~\cite{Jel1} and ceramics~\cite{Dmi1}, suggesting that the
region where the dissipation is dominated by the nucleation of PSC
is not restricted to the vicinity of $T_c$. In addition, these
experiments would imply, if the PSC interpretation is correct,
that the PSC picture, developed for a 1D situation, is still valid
in a 2D superconductor.

Nevertheless only few publications interpret the steps in the I-V
curves at low temperature in terms of PSC. Discontinuities in the
I-V characteristics are generally attributed to heating effects
(or 'bolometric' effects) because thermal instabilities can
produce very fast voltage variations in thin films. Indeed the
thermal response time, which is equal to the heat capacity of the
film divided by the thermal conductance of the film-to-substrate
boundary, is typically a few nanoseconds.

Some reports however do not interpret the voltage jumps induced by
the application of a high current as hot spots, but the proposed
explanations are all different from the PSC nucleation
scenario.~\cite{Sto1,Asi1,Xia1,Doe1,Xia6}

At low temperature, the steps observed in Nd$_{2-x}$Ce$_x$CuO$_y$
were explained in terms of a strongly energy dependent density of
states~\cite{Sto1}. In this picture, unlike the PSC model, these
steps do not reflect the nucleation of a spatially limited domain
surrounded by low dissipative regions, but are caused by an
oscillation of the density of state as a function of energy in the
superconducting mixed state. This model is valid only in the clean
or superclean limit and therefore cannot be a general model for
all superconductors. In contrast, steps attributed to {\it
spatially limited domains} were reported in narrow
YBa$_2$Cu$_3$O$_{7-\delta}$ bridges and interpreted as an effect
of edge pinning~\cite{Asi1}.

Close to $T_c$, the application of a high current density has been
seen to induce large voltage jumps bringing the sample close to
the normal state resistance in
YBa$_2$Cu$_3$O$_{7-\delta}$~\cite{Xia1,Doe1} and in
Bi$_2$Sr$_2$CaCu$_2$O$_{8+\delta}$~\cite{Xia6} thin films. These
discontinuities in the I-V characteristics were attributed to flux
flow instabilities, an effect predicted 25 years ago by Larkin and
Ovchinnikov~\cite{Lar1}, involving the density of states in the
vortex core.

It is likely that a single mechanism is at the origin of most of
the abrupt voltage variations observed in high $T_c$
superconducting films at high current density. In an effort to
provide an overall picture of this extreme regime, we have
performed time resolved transport measurements on two different
compounds, Bi$_2$Sr$_2$CaCu$_2$O$_{8+\delta}$ (BSCCO) and
Nd$_{1+x}$Ba$_{2-x}$Cu$_3$O$_{7-\delta}$ (NBCO) thin films, over a
wide temperature range.

Based on our measurements, we show that a very systematic and
general phenomenon appears in HTS compounds at a certain critical
current $j^*(T)$, causing a rapid development of highly
dissipative domains. We argue that, unlike what is concluded for
conventional superconductors, even at low temperature these
domains do not result from a thermal instability. Instead they
behave just like 1D PSC's observed in metallic superconductors
near $T_c$.

We also show that the current induced breakdown of
superconductivity seems to result from a large scale propagation
starting with an elementary PSC. A full understanding of the way
the current induces a switch into the normal state is important in
the development of certain applications, in particular in the use
of HTS thin films as fault current limiters~\cite{Dec1},
bolometric detector~\cite{Kre1} or ultrafast superconductive
switches~\cite{Ger1}.


\section{experimental details}


The samples used for the present study are BSCCO and NBCO
epitaxial films, 200 nm thick, deposited by rf magnetron
sputtering on (100) oriented MgO substrates. The NBCO film are
Nd-rich (the nominal target composition is
Nd$_{1.15}$Ba$_{1.85}$Cu$_3$O$_{7-\delta}$). This prevents the
formation of screw dislocations, resulting in good
cristallinity~\cite{Sal2}, but $T_c$ decreases
slightly~\cite{Kra2}. MgO was chosen for its high thermal
conductivity (about 25 times larger than SrTiO$_3$ and LaAlO$_3$)
in order to optimize the heat removal during the measurements. As
revealed by x-ray analysis, all films were $c$-axis oriented. 60
$\mu$m wide bridges were patterned by standard photolitography
into a four-probes configuration, with a distance of 500 $\mu$m
between the voltage contacts. A gold layer was deposited on all
the contact pads to reduce their resistance. The critical
temperatures after the patterning, taken at mid-transition, are 80
K for the NBCO and 76 K for the BSCCO thin film.

To limit the Joule heating, the current is applied by single
pulses. The purpose-built pulsed current source could deliver
square or triangular pulses with an amplitude up to 1 Amp and a
duration ranging from 10 $\mu$s to 1 s. During square pulses, the
current varied by less than 1$\%$. Current stability is crucial
since any spike above the critical value can result in an abrupt
resistance change, leading to wrong interpretations. The voltage
is recorded as a function of time by a Tektronics digital
oscilloscope. The current is measured simultaneously on the second
channel of the oscilloscope by picking the voltage across a 1
$\Omega$ resistor in series with the sample.

Resistive measurements were performed from 4 K up to $T_c$. The
sample is glued by a GE heat conducting varnish onto a copper
block and surrounded by 100 mbar of helium gas. The sample
temperature is controlled by a heater and monitored by a
carbon-glass probe placed inside the copper block. Once a pulse is
applied, the heat front crosses the substrate and reaches the
copper block in about 40 $\mu$s. In the dissipative state (normal
or mixed state), the heat produced in the film causes a smooth
resistance increase, consistent with a simple simulation of heat
diffusion through the substrate to the copper block. The
thermometer probe is however too far to record the real time
temperature increase of the film. Therefore we use the evolution
of the film resistance to obtain information about its temperature
{\it during a pulse}. A magnetic field generated by a
superconducting magnet in a persistent mode can be applied
parallel to the $c$-axis.


\section{results}


\subsection{critical current}

The currents applied in this experiment are well above the
critical current which is traditionally defined at the onset of
resistance. At high current density (above 10$^6$ A/cm$^2$), the
measured voltage across the bridge exhibits abrupt variations as a
function of current or temperature. Two kinds of voltage jumps are
observed: small steps where the resistance increase is less than 1
\% of $\rho_n$ ($\rho_n$ being the normal resistivity just above
$T_c$), and large jumps where the resistance reaches a value close
to $\rho_n$. The voltage increase during a small jump is
instantaneous at our time resolution, i.e. the voltage variation
is faster than the microsecond rise time of our differential
amplifier. With constant current pulses, we show in this section
that both kinds of voltage steps are due to a single mechanism,
which occurs when the current reaches a critical value that we
call $j^*(T)$.

\begin{figure}
{\centerline{\large Figure 1}}
\centerline{\epsfig{figure=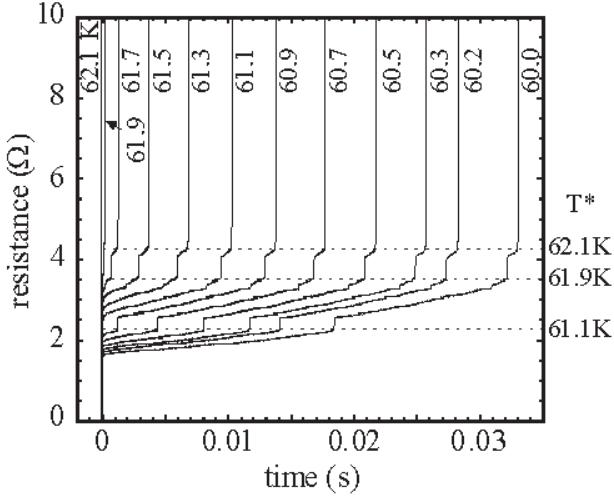,width=8.5cm} } \caption{Resistive
response of the NBCO film when a constant current pulse of 1.9
10$^6$ A/cm$^2$ is applied ($B=0.1$ T). For each curve, the
initial temperature indicated on the top of the figure is changed;
however, each step occurs always at the same temperature $T^*$
(displayed on the right side). The sample temperature at a certain
resistance is determined by varying the initial temperature until
this resistance appears immediately at the beginning of the pulse.
} {\label{f:sameR_steps}}
 \vspace{2cm}
\end{figure}

A typical set of results for identical square pulses is shown in
fig.~\ref{f:sameR_steps} for the NBCO thin film at 0.1 T. The
applied current density  is 1.9 10$^6$ A/cm$^2$ and each curve
corresponds to a different sample temperature, measured just
before the pulse application, varying from 60.0 K to 62.1 K. Each
curve exhibits two clear steps before the total breakdown of
superconductivity occurs, with the resistance rising up to
$\rho_n$. Note that each particular jump occurs at a given
resistance independent of the initial temperature.

The smooth increase between the steps is due to Joule heating but
cannot be recorded in real time by the temperature probe, located
in the sample holder. The probe will indicate a temperature
corresponding only to the initial resistance appearing at the
beginning of the pulse. We use these initial values to obtain the
$R$ versus $T$ function {\it at a particular current value}.
Knowing this calibration function we can relate, at any time
during the pulse, the resistance to the sample temperature, and in
particular we can determine the temperature at which each jump
occurs.

We observe that for a given current, each particular jump (small
or large) always occurs at the same temperature, that we call
$T^*$, indicated on the right hand side of
fig.~\ref{f:sameR_steps}. $T^*$ depends on the current ($T^*(j)$
decreases when the current increases), and by varying $j$, we
observe the steps over a wide temperature range, from 4 K up to a
temperature close to $T_c$ (0.7 $T_c$), in both NBCO and BSCCO
films. Above a certain temperature, which turns out to be near
$T_c$, the uniform resistance presumably caused by vortex motion
is so high that the sample temperature 'runs away' and the abrupt
transition is replaced by a smooth upturn of the voltage as a
function of time.

\begin{figure}
{\centerline{\large Figure 2}}
\centerline{\epsfig{figure=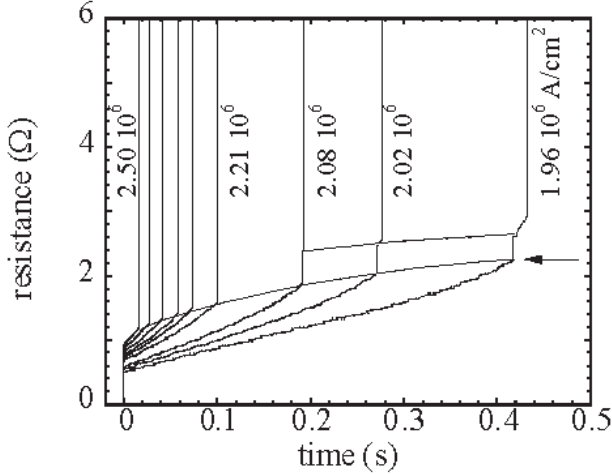,width=8.5cm} } \caption{Resistive
response of the NBCO film to square current pulses with different
amplitudes ($B=0.1$ T). The initial temperature before the pulse
application is 55 K. When decreasing the current amplitude, the
small resistance jump indicated by the arrows turns into a large
jump where the total sample resistivity reaches the normal state
resistance. } {\label{f:identity}}
 \vspace{2cm}
\end{figure}

The data of fig.~\ref{f:sameR_steps} indicate that in one sample
for a given current density, there are several temperatures $T^*$,
each one corresponding to a different location along the bridge.
The local character of the dissipation was established by adding a
third voltage contact in the middle of the bridge: within this
configuration, when measuring simultaneously the voltage on both
sides of the bridge, the small jumps were detected on only one
segment, never on both. As a consequence, while the large jump
corresponds to a transition of the whole bridge, each small step
is produced by the sudden formation of a localized highly
dissipative domain (HDD). Fig.~\ref{f:sameR_steps}
and~\ref{f:identity} show that these domains can persist over
relatively long times without affecting the superconducting state
in the rest of the bridge. The extreme reproducibility of the HDD
formation (even when the films were warmed up to room temperature
and cooled down again) suggests that the $T^*$ spreading for a
given current is due to slight spatial inhomogeneities along the
bridge.

One of the crucial points, demonstrated in fig.~\ref{f:identity},
is that the formation of an HDD is also the origin of the global
sample transition characterized by the large voltage jump. In this
experiment performed on the NBCO thin film, each curve was
measured with a different current (explaining the variation of the
jump temperature $T^*$). Above a certain current, the small step
indicated by the arrow is replaced by a global transition.

\begin{figure}
{\centerline{\large Figure 3}}
\centerline{\epsfig{figure=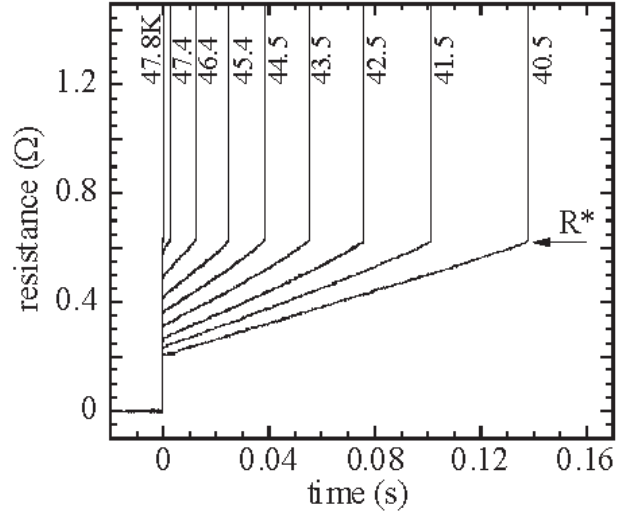,width=8.5cm} } \caption{Resistive
response of the NBCO film to pulses of 3.1 10$^6$ A/cm$^2$ at 0.1
T. The sample temperature before the pulse application is
indicated on the top of each curve. This temperature is increased
from 40.5 K to 47.8 K. At the latter temperature the transition
occurs immediately at the beginning of the pulse. 47.8 K is
therefore interpreted as the temperature $T^*$ corresponding to
the applied current.} {\label{f:sameR}} \vspace{10cm}
\end{figure}

We now move to the shape of $T^*$ in the $j-T$ plane. To obtain
experimentally the reciprocal function $j^*(T)$, we have to detect
the formation of a particular HDD over the whole temperature
range. The easiest way is to look at the large voltage jump.
Fig.~\ref{f:sameR} illustrates the procedure to obtain one point
of $j^*(T)$: we fix the current, here 3.1 10$^6$ A/cm$^2$, and
then we increase the initial sample temperature, until the global
transition occurs immediately at the beginning of the pulse. In
this example the sample switches at 47.8 K, i.e. $j^*(47.8 K)=3.1$
10$^6$ A/cm$^2$. By applying the same procedure for small steps,
we have verified that the critical current associated with the
small steps has the same temperature dependence than the large
jumps on some limited intervals (we could not follow a particular
small jump from 4 K up to $T_c$). This confirms the conclusion
drawn from fig.~\ref{f:identity}, namely that the small and the
large steps have the same origin.

\begin{figure}
{\centerline{\large Figure
4}}\centerline{\epsfig{figure=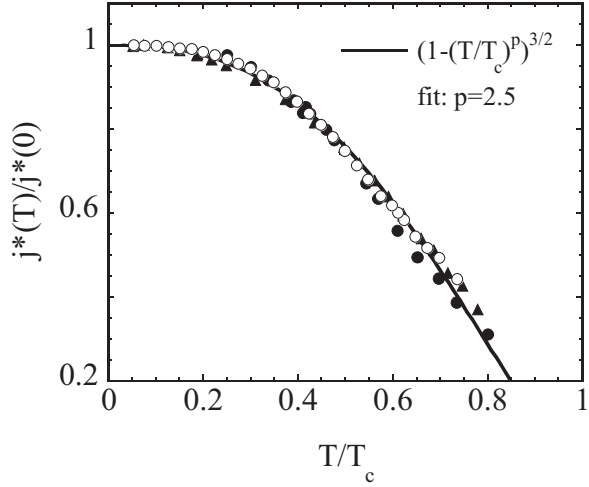,width=8.5cm} }
\caption{Temperature dependence of $j^*$ normalized at
zero-temperature. It represents the nucleation current of a highly
dissipative domain. Similar behavior is exhibited by BSCCO
(triangles)  and NBCO (filled circle) at 0.1 T, as well as NBCO at
1 T (empty circle). The solid line is a fit of the BSCCO data with
$(1-(T/T_c)^p)^{3/2}$, where $p$ is the fitting
parameter.}{\label{f:phasediag}}
 \vspace{2cm}
\end{figure}

Fig.~\ref{f:phasediag} shows the results for NBCO (filled circles)
and BSCCO (filled triangles) films. It is remarkable that for both
compounds the temperature dependence of $j^*$ is the same.
$j^*(0)$, the extrapolated value at $T = 0$, which represents the
highest current density that can be carried by the film while
remaining in the superconducting state, is also found to be
similar in the two compounds. $j^*(0)$ in the NBCO film was in
fact slightly larger than in BSCCO (5.8 10$^6$ A/cm$^2$ instead of
4.4 10$^6$ A/cm$^2$), possibly due to the higher $T_c$ of the
former.

It should be noted that upon increasing the magnetic field
$j^*(T)$ is globally reduced but the shape of the $j^*$ versus $T$
curve remains the same (see fig.~\ref{f:phasediag} for the 1 T
case). The zero-field data are not shown here because they do not
exhibit significant differences compared with the 0.1 T case.

\begin{figure}
{\centerline{\large Figure 5}}
\centerline{\epsfig{figure=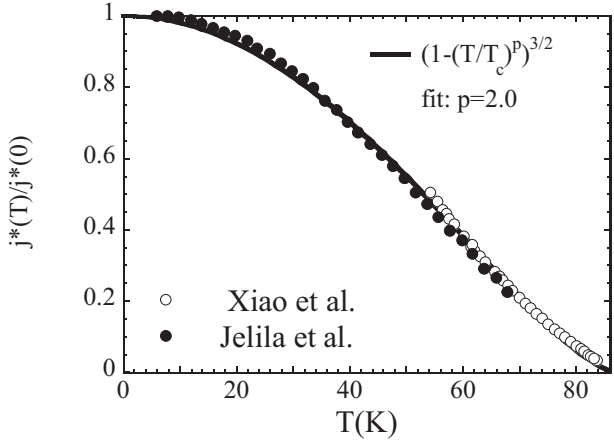,width=8.5cm} }\caption{Normalized
critical current producing voltage jumps in
YBa$_2$Cu$_3$O$_{7-\delta}$ thin films measured by Jelila et
al.~\cite{Jel1} and by Xiao et al.~\cite{Xia1}. Both data are
fitted by a single curve with two fitting parameters: $p$ and
$T_c$. The fit gives: $p=2.0$ and $T_c=87.2$ K.}
{\label{f:phasediagother}}
 \vspace{2cm}
\end{figure}

Hence, the temperature dependence of $j^*$ appears to be a general
characteristics of HTS compounds. It can be well fitted by:

\begin{equation}\label{eq:j*(T)}
  \frac{j^*(T)}{j^*(0)}=\left( 1-\left( \frac{T}{T_c} \right)^p \right)^{3/2},
\end{equation}

with $p$ ranging from 2 to 2.5. It is worth noting that the
temperature dependence described by Eq.~\ref{eq:j*(T)} also fits
other reported measurements of high current-induced phenomena.
Fig.~\ref{f:phasediagother} displays the critical current for PSC
formation found by Jelila et al.~\cite{Jel1} and the critical
current giving rise to a voltage instability near $T_c$ measured
by Xiao et al.~\cite{Xia1}, both in YBa$_2$Cu$_3$O$_{7-\delta}$
films. Like for our measurments, the temperature dependence of
these data agrees with expression~\ref{eq:j*(T)}.

The picture that emerges so far is that a HDD nucleates somewhere
along the bridge when the temperature reaches a critical value
$T^*(j)$. In some cases, it rapidly propagates to the whole
sample, causing the complete switching into the normal state. The
propagation speed of a highly dissipative region at high current
densities has been recently measured in large
YBa$_2$Cu$_3$O$_{7-\delta}$ films at 77 K \ \cite{Dec2}: it ranges
from 20 to 120 m/s. In our experiment, because of the shortness of
the bridge, the voltage variation during the propagation seems
instantaneous. In conclusion we find that the small steps and the
total transition have a common origin. To understand this origin
we will focus, in the next section, on small voltage jumps.

\subsection{I-V characteristics}
\label{s:IVC}

To investigate the nature of the HDD, we present here the I-V
characteristics of these domains at low temperature. The aim is to
see the HDD signature by sweeping the current, instead of the
temperature as in the previous section, across the critical value
$j^*$ associated with one particular domain. In this study, we
chose to look at the BSCCO film, where numerous small steps could
be observed.

\begin{figure}
{\centerline{\large Figure
6}}\centerline{\epsfig{figure=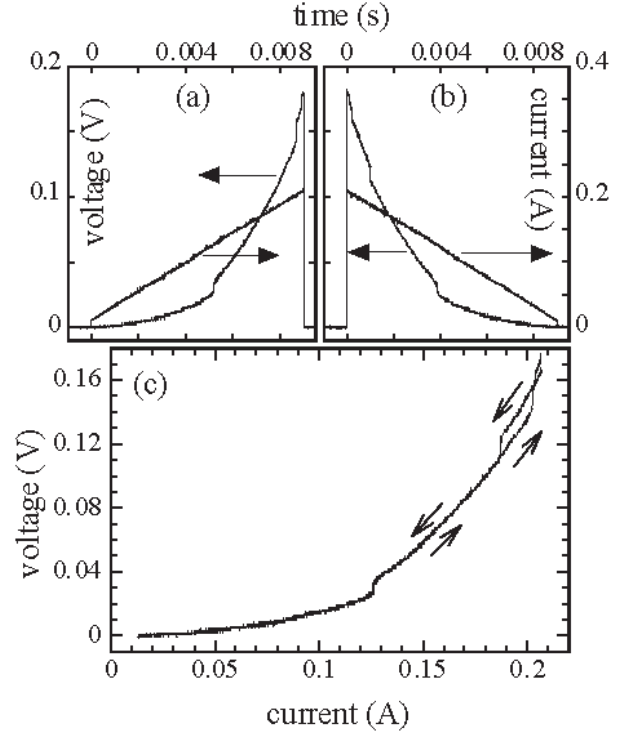,width=8.5cm} } \caption{Voltage
response of the BSCCO film at 30 K to an increasing (a) and
decreasing (b) current ramp of 20 A/s. The deduced IV
characteristics (c), shows the good matching between the two
curves and the hysteresis on the HDD formation.}
{\label{f:IV_demo}}
 \vspace{1cm}
\end{figure}

In the following experiment the voltage response is recorded
during a current ramp. Unlike the experiment using square pulses,
we need that the temperature does not increase irreversibly during
the application of the current, i.e that the measured voltage is
history-independent (except for the possible hysteresis in the HDD
formation and disappearance). This absence of heating was verified
by applying upward and downward current ramps (see
fig.~\ref{f:IV_demo} (a) and (b)). The ramp duration was reduced
so that both curves matches together (see fig.~\ref{f:IV_demo}
(c)). The only difference is the value of $i^*$ for the second
step. Alternatively, we observed that the I-V curves remain the
same when the ramping rate is varied. For the example of
fig.~\ref{f:IV_demo}, the same characteristics are found when the
ramping rate is raised from 20 A/s to 500 A/s.

Fig.~\ref{f:IV} shows that the raw I-V characteristics (curve A)
are different below and above $i^*_1$, the critical current
corresponding to the first step. Below $i^*_1$, the I-V curve can
be well fitted by a power law: $U \sim i^q$ and represents the
homogeneous dissipation in the whole bridge due to vortex motion
(curve B represents the extrapolation of this voltage to high
currents). Above $i^*_1$, the voltage comes from the uniform
dissipation of vortices (curve B) and the contribution of the HDD.
Therefore, to obtain the voltage of a {\it single} HDD, we have to
subtract curve B from the raw data.

\begin{figure}
{\centerline{\large Figure 8}}
\centerline{\epsfig{figure=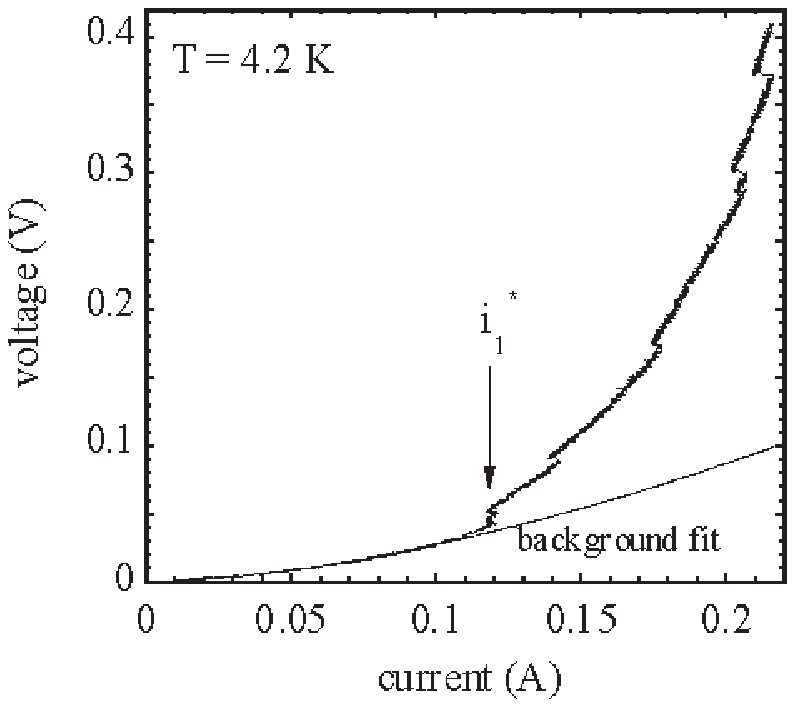,width=8.5cm} }\caption{I-V
characteristics of the BSCCO film at 4.2 K and 0.1 T. For each
critical current the jump occurs with a slope given by the source
impedance (here 1.5 $\Omega$). The differential resistance of the
linear portions increase after each jump.} {\label{f:PSC}}
 \vspace{1cm}
\end{figure}

The resulting I-V characteristics (curve C) are non-ohmic: the I-V
have piecewise linearity but the extrapolated zero-voltage current
is non-zero. The differential resistance of the first domain can
however be estimated: $dU/di=0.6$ $\Omega$. Other domains are
successively created when the current is raised (see
fig.~\ref{f:PSC}). A very noticeable feature is that after each
jump, the slope of the linear portion increases.

It should be noted that the characteristics displayed in
fig.~\ref{f:IV}, performed at 30 K, are representative of the
BSCCO behaviour at low temperature, because we do not see a
significant change below 30 K. For instance, as one can see in
fig.~\ref{f:phasediag}, $j^*(T)$ does not vary significantly at
low temperature.


\section{discussion}


Since in metallic superconductors, resistive domains result, at
low temperature, from thermal instabilities, we will first examine
if the voltage steps can be attributed to hot spots creation. By
"hot spot" we mean a localized domain with a temperature {\it
above $T_c$}. A hot spot can result from a thermal instability
that takes place when the heat generated in the system (which can
be the film) cannot be evacuated into the heat sink. Such an
instability produces necessarily domains above $T_c$ \
\cite{therm_instab}.

\begin{figure}
{\centerline{\large Figure 7}}
\centerline{\epsfig{figure=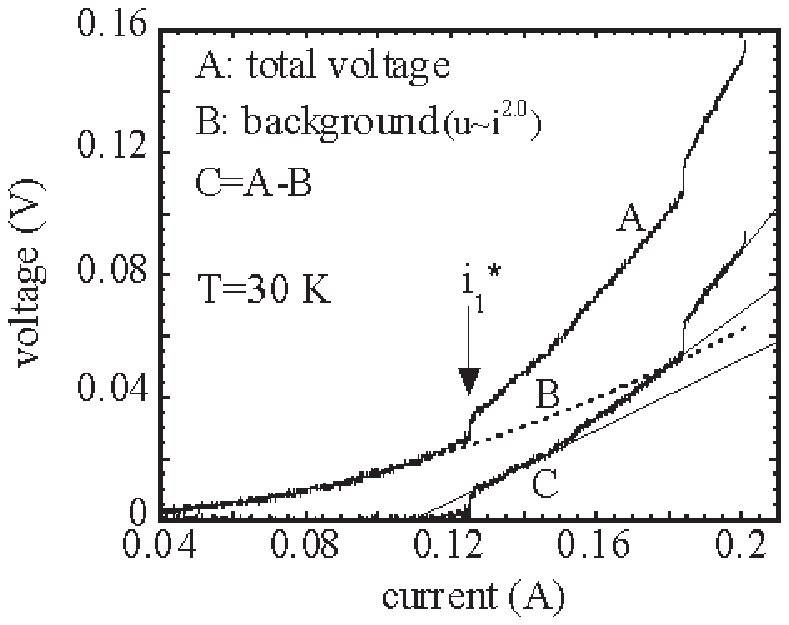,width=8.5cm} }\caption{I-V
characteristics of BSCCO at 30 K and 0.1 T showing the first steps
obtained by applying a triangular current pulse (curve A). Apart
from some hysteresis on HDD's formation, the same voltage is found
by increasing or decreasing the current. Before the first step,
the dissipation is homogeneous and can be fitted by the curve B,
with an exponent $q\approx 2.0$. By subtracting this contribution
one obtains the I-V characteristics of HDD's only (curve C).}
{\label{f:IV}} \vspace{1cm}
\end{figure}

\begin{figure}
{\centerline{\large Figure 9}}
\centerline{\epsfig{figure=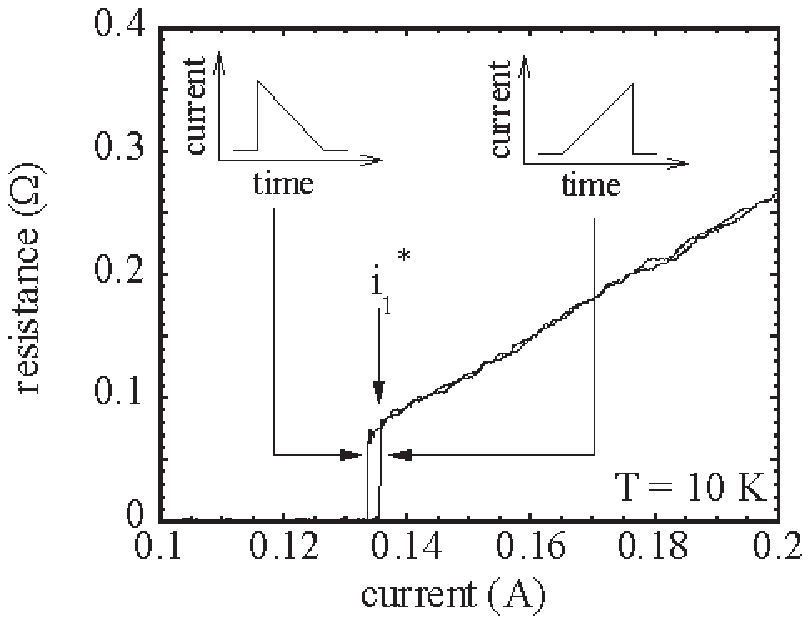,width=8.5cm} }\caption{Resistance
of the first HDD defined as the voltage divided by the total
current in the BSCCO film at 10 K  and 0.1 T. The current is
ramped at the rate of 30 Amp/s, so that the whole curve is taken
in 3 ms. Above $i_1^*$, there is a good agreement between the
curves obtained with the upward and the downward ramp. As observed
for many HDD's, the current induced nucleation is hysteretic.}
{\label{f:Ri}}
 \vspace{2cm}
\end{figure}

In Fig.~\ref{f:Ri} we plot the resistance of a single HDD as a
function of the current, at $T=10$ K. Here, the resistance is
defined as the voltage divided by the total current, which is the
appropriate quantity for a normal domain. The HDD resistance is
obtained after subtracting the background, as in fig.~\ref{f:IV},
and therefore it is the resistance we would obtain by placing
local voltage probes just across the HDD.

If the HDD is a hot spot, then its length can be deduced directly
from the ratio of its resistance over the total bridge resistance
just above $T_c$, i.e 31 $\Omega$. This yields a length of 1.3
$\mu$m. As a consequence, right after the nucleation the HDD would
be a thin line (1.3 x 60 $\mu$m) cutting the bridge
perpendicularly to the current direction.

The main point comes directly from the inspection of
fig.~\ref{f:Ri}. Since the HDD, when created, does not have an
ohmic behavior, it cannot be a normal zone above $T_c$, unless the
domain {\it temperature} or {\it size} changes as a function of
the applied current. However, above $i^*_1$, the I-V curve are the
same with an ascending or a descending current ramp. This
indicates that the application of the current does not result into
an irreversible heating or expansion of the HDD. Therefore, if
these changes occurs, they do in a {\it stationnary} manner.

Gurevich and Mints~\cite{Gur1} have shown that, in a stationary
situation, the hot spot size must {\it decrease} with an
increasing current, in order to evacuate the excess of heat
generated in the domain. Hence, a change in the domain size cannot
explain the resistance increase of fig.~\ref{f:Ri}.

Furthermore, the resistance increase is too large to be explained
by a temperature increase alone. Indeed, if the value just after
the nucleation ($R_{nucl}=0.07$ $\Omega$) corresponds to the
normal state at 90 K, then when the resistance reaches 0.26
$\Omega$ (on the right side of fig.~\ref{f:Ri}), the HDD would be
over room temperature (334 K). Furthermore, in this sample, I-V
measurement were performed up to 0.3 Amp, and the resistance
continued to increase, corresponding to a hypothetical hot spot of
700 K. At these temperature, the HDD would irreversibly damage the
film, in particular causing a loss of oxygen, and the I-V would
not be reproducible.

Thus, the hot spot model cannot explain the observed voltage
steps. This conclusion is also supported by the following
quantitative argument.

Let us estimate the temperature increase $\Delta T$ in a
hypothetical hot spot of 1.3 $\mu$m, just after its creation, when
we neglect the heatlink with the helium gas. Since all the power
produced in excess must be evacuated toward the substrate, we
have: $R_{nucl} i^{*2}_1 = S \lambda \Delta T$, where $S$ is the
area of the interface with the substrate and $\lambda$ the thermal
conductance between the film and the substrate. Taking $\lambda =
10^7$ WK$^{-1}$m$^{-2}$, a value generally obtained for the
boundary conductance between YBa$_2$Cu$_3$O$_{7-\delta}$ and
MgO~\cite{Nah1,Car1,Ser1}, the temperature increase in the HDD
right after its formation would be around 2 K. This rough estimate
yields that the power generated in the HDD is not sufficient to
maintain the temperature above $T_c$.

Plotted as the voltage versus current, as in fig.~\ref{f:IV}, we
see that the only deviation from an ohmic behaviour is a non-zero
intersect at zero voltage. Therefore the HDD can be understood as
a normal region where the current that generates dissipation is
just a fraction of the total current. In other words, in addition
to the normal current, a supercurrent would be flowing across the
HDD.

In fact, with respect to the three features, piecewise linearity,
positive zero-voltage intersect and hysteresis, the I-V curves
obtained here resemble the measurements performed in tin
whiskers~\cite{Mey1} and microbridges~\cite{Sko1}. In both
systems, dissipation was well explained by the standard model of
PSC nucleation of Scokpol, Beasley and Tinkham (SBT)~\cite{Sko1}.
According to this model, only part of the total current
contributes to the dissipation of a PSC and this current produces
a normal state resistance. Therefore, the I-V characteristics of a
single PSC is given by:

\begin{equation}\label{eq:IVPS}
  V=2 \Lambda \rho (j-j_0),
\end{equation}

where $\rho$ is the normal resistivity, $\Lambda$ the diffusion
length of the quasiparticle generated in the PSC core, and $j_0$
the time average supercurrent. In the SBT description, the
suppression of the order parameter occurs in a small region of
size $\xi$, but the measured voltage drop is governed by the
diffusion of quasiparticles over a much larger distance $\Lambda$.

Several experiments carried out in Sn films show that the "PSC
kind" of I-V, i.e. a series of linear section separated by sharp
steps is also visible in wide samples~\cite{Lli1,Kon1,Dmi2}. It
suggests that a 2D system exhibits the same behaviour as a narrow
superconductor, but the 1D core is replaced by a thin line
crossing the bridge width (i.e. a phase-slip line) or a channel of
rapidly moving vortices~\cite{Web1}. This extension to a 2D case
is also supported by theory~\cite{Web1,And2}.

Therefore, we can evaluate $\Lambda$, the extension of the HDD,
using Eq.~\ref{eq:IVPS}: the differential resistance of a single
HDD is $2 \Lambda \rho /\sigma$, where $\sigma$ is the bridge
cross section. This determination requires an estimate of the
normal carriers resistivity $\rho$ at a temperature $T<T_c$.
Taking into account the reduction of the scattering rate at low
temperature, one can assume a resistivity of the form
$\rho(T)=\rho_d+\rho_s(T)$ where $\rho_d$ is due to defects and
becomes dominant at low temperature~\cite{Fli2} and $\rho_s(T)$ is
the linear component measured above $T_c$. Qualitatively such a
behavior for the normal state resistivity is observed under high
magnetic fields~\cite{Ono1}. Since we see little change below 30 K
in the I-V characteristics, we estimate $\rho_d$ as $\rho_n
T_d/T_{onset}$ where $T_d$ is the saturation temperature we
estimate to be $30$ K and $T_{onset}$ is 80 K , the temperature
just above the transition. This estimate yields 30 $\mu \Omega$cm
for the resistivity entering in expression~(\ref{eq:IVPS}).
Therefore, the first HDD shown in fig.~\ref{f:IV}, with
differential resistance of 0.6 $\Omega$, has a characteristic
length $2 \Lambda = 27\pm 10$ $\mu$m, much larger than in the hot
spot case. The error allows for the large uncertainty in $\rho$.
Note that the first HDD could include several PSC and therefore
this estimate represents an upper limit of the PSC size.

Coming back to the temperature dependence of $j^*$, it is
interesting to note that such a temperature dependence is what is
most likely expected for the {\it depairing current} $j_d(T)$ in
high $T_c$ superconductors. Indeed, using Ginzburg-Landau
relations for $j_d$, we have:

\begin{equation}\label{eq:jd}
 j_d(T)\propto \frac{1}{\lambda^2(T)} \frac{1}{\xi(T)} \propto \left( \frac{1}{\lambda^2} \right)^{3/2},
\end{equation}

where the $1/\lambda^2(T)$ factor is the temperature dependence
coming from the superfluid density and $1/\xi(T)$ is the
contribution of the critical velocity. The last proprotionality is
obtained by assuming that the Ginzburg-Landau parameter
$\kappa=\lambda/\xi$ is only weakly temperature dependent. The
temperature dependence of $j_d$ will be the one displayed on the
right hand side of eq.~\ref{eq:j*(T)}, because the temperature
dependence of the superfluid density measured from zero to $T_c$
is well described by $1-(T/T_c)^p$ with $p$ ranging from 1.7 to
2.9~\cite{Ple1,Anl1,Ris1}.

However, the amplitude of $j^*$ is well below the depairing
current (about two orders of magnitude). In a perfect
superconducting filament, in contrast, a PSC would develop when
the supercarriers reaches the critical velocity. The fact that a
HDD appears at a current representing a small fraction of $j_d$
implies that the energy cost to create a phase-slip line or a
channel of fast moving vortices is well below the one needed to
destroy completely superconductivity in the film. A theoretical
prediction of $j^*$ would require an understanding of the exact
mechanism causing the development of the PSC. This is presently
beyond the scope of this article. Nevertheless, we can deduce from
the similarity of $j^*$ in the different compounds that this
mechanism is not dependent on the nature of the defects, and
therefore not related to vortex pinning.


\section{conclusion}


In every high $T_c$ thin film of BSCCO and NBCO we measured, the
high current part of the I-V characteristics exhibits some sharp
voltage variations. These are robust features and are observed
over a wide temperature range, from 4 K up to the vicinity of
$T_c$. This behavior of superconducting films traversed by a high
density of current seems very systematic, since it has been
reported by other groups in
YBa$_2$Cu$_3$O$_{7-\delta}$~\cite{Jel1,Asi1,Xia1,Doe1} or
Nd$_{2-x}$Ce$_x$CuO$_y$~\cite{Sto1}, with various interpretations.
We believe that a very general mechanism occurs in the
superconducting state at high current density independently of the
defects, such as pinning centers, etc. By means of time resolved
resistance measurements, we have shown that this mechanism
accounts for both small voltage steps and a more dramatic event
like the current-induced breakdown of superconductivity. The small
voltage steps are associated with the nucleation of local
dissipative domains. We have brought new arguments showing that
these domains do not result from thermal instabilities, like hot
spots (the situation is quite different in bulk superconductors,
where Joule heating may dominate). Another mechanism, directly
related to superconductivity must therefore be at the origin of
the formation of these domains. Even though the samples cannot be
considered as one-dimensional, the transport properties are in
good agreement with a dissipative state dominated by phase-slip
centers, in which the superconductor is locally driven far from
its equilibrium point. If this interpretation is correct,
transport measurements at high current density allow to access
intrinsic properties of the superconductor. The fundamental
character of this dissipation process is confirmed by the
temperature dependence of the critical current which is similar in
different high $T_c$ compounds and turns out to be close to the
one expected for the depairing current.

\acknowledgements

We thank P. Martinoli for stimulating discussion and careful
reading of the manuscript. This work was supported by the Swiss
National Science Foundation.


\bibliographystyle{prsty}


\begin{thebibliography}{10}
\bibitem{therm_instab}
The instability occurs when the slope of $R(T)$ exceeds some
critical value, and the system stabilizes when the slope
decreases, which is only possible above the NS transition. For a
complete discussion of bistabilities, see ref \ \cite{Gur1}.
\end{thebibliography}



\newpage

\end{multicols}
\end{document}